\documentclass[12pt]{iopart}

\expandafter\let\csname equation*\endcsname\relax

\expandafter\let\csname endequation*\endcsname\relax

\usepackage{graphicx}
\usepackage{amsthm, amsmath, amssymb}
\usepackage{dsfont}
\usepackage{bm}
\usepackage{tikz}
\usepackage{tikz-3dplot}
\usepackage{tkz-euclide}
\usetikzlibrary{automata,positioning}
\usepackage{ctable}
\usepackage{array}
\newcolumntype{?}{!{\vrule width 1pt}}
\usepackage{epstopdf}
\usepackage[hyperindex,breaklinks,unicode,colorlinks=true,linkcolor=blue,citecolor=blue,urlcolor=blue]{hyperref}
\usepackage[all]{hypcap} 
\usepackage{xcolor}

\begin{document}

\title[Order-Disorder Transition in Delay Vicsek Model]{Order-Disorder Transition and Phase Separation in Delay Vicsek Model}

\author{Robert Horton$^{1,\dagger}$, Viktor Holubec$^{2,\star}$}

\address{$^1$ Institut f\"ur Theoretische Physik, Universit\"at Leipzig,  Postfach 100 920, D-04009 Leipzig, Germany}

\address{$^2$ Charles University,  Faculty of Mathematics and Physics, 
 Department of Macromolecular Physics, V Hole{\v s}ovi{\v c}k{\' a}ch 2, 
 CZ-180~00~Praha, Czech Republic}

\ead{$\dagger$ robert@rodmell.org}
\ead{$\star$ Corresponding author, viktor.holubec@mff.cuni.cz}

\vspace{10pt}
\begin{indented}
\item[]\today
\end{indented}

\begin{abstract}
Interactions in active matter systems inherently involve delays due to information processing and actuation lags. We numerically investigate the impact of such delays on the phase behavior of the Vicsek model for motile active matter \textcolor{black}{at a large but fixed system size}. While the delayed Vicsek model retains the same three phases as the standard version—an ordered state, a liquid–gas coexistence state, and a disordered state—the presence of delay qualitatively alters the system’s dynamics. At \textcolor{black}{the relatively} high velocity considered in this study, the critical noise for the transition between the ordered and coexistence states exhibits a non-monotonic dependence on delay, whereas the critical noise required for the transition to the disordered state increases with delay. Consequently, the width of the noise interval in which phase separation occurs broadens with increasing delay. Short delays stabilize the ordered phase, while long delays destabilize it in favor of the coexistence phase, which is consistently stabilized compared to the disordered state. Furthermore, the number of bands observed in the coexistence state at a given noise increases, and the time required for their formation decreases with delay. This acceleration is attributed to the emergence of swirling structures whose typical radius grows with increasing delay. Our results demonstrate that time delay in the Vicsek model acts as an effective control parameter for tuning the system’s dynamic phase behavior.
\end{abstract}

%

%
\submitto{\NJP}

\section{Introduction}

Active matter physics~\cite{Vicsek201271,ramaswamy2010mechanics,bechinger2016active} applies statistical physics tools to uncover the key ingredients underlying highly organized many-body dynamics in systems composed of nonequilibrium agents, such as bacterial colonies~\cite{zhang2010collective,ben1994generic}, bird flocks~\cite{cavagna2018physics}, human crowds~\cite{shahhoseini2018pedestrian,Helbing2002SimulationOP}, and robotic swarms~\cite{mijalkov2016engineering,vasarhelyi_optimized_2018,khadka2018active}. Interactions in these systems are often governed by a perception-reaction feedback loop, leading to deviations from standard physical symmetries, such as reciprocity~\cite{Fruchart2021,loos2020irreversibility}, and introducing time delays relative to the stimuli~\cite{scholl2009time,loos2021stochastic}. While the role of reciprocity has been extensively studied in recent years, the effects of time delay remain comparatively less explored.

One of the best-known effects of delayed interactions in active matter systems is the phenomenon of phantom traffic jams, which arise due to the delayed reactions of drivers~\cite{Davis2003}. Beyond inducing oscillatory behavior—a characteristic also observed in inertial motion—time-delayed interactions can lead to multistability, instabilities, and even chaotic dynamics~\cite{forgoston2008delay,piwowarczyk_influence_2019,geiss_brownian_2019,loos2019heat}. Recent experiments~\cite{khadka2018active,Munos2021,wang_spontaneous_2023} with feedback-driven, overdamped artificial microswimmers~\cite{Franzl2021} have further demonstrated that delay-induced aiming errors lead to chirality~\cite{wang_spontaneous_2023} and revealed the existence of an optimal transport speed~\cite{Munos2021}, similar to the optimal run-and-tumble times observed in bacteria~\cite{Romanczuk2015,Diz-Munoz2016}.

In active matter models, delayed repulsion from neighboring agents~\cite{Lowen2019} and even from an agent’s own past position~\cite{Kopp2023} can induce activity, while delayed attraction was shown to induce flocking, swarming, and swirling motions~\cite{forgoston2008delay}. In Vicsek-type models with alignment interactions, short delays enhance order, whereas long delays can disrupt it~\cite{Erban2016, piwowarczyk_influence_2019, holubec_finite-size_2021}. Besides, simulations suggest that time delays change the nature of information transfer in the Vicsek model (VM)  from diffusive to ballistic~\cite{geis_signal_2022}.
Introducing repulsion into the delayed VM leads to the emergence of mixed regions of order and disorder~\cite{Pakpour2024}.
Moreover, delays alter the dynamical finite-size scaling exponent governing space-time correlations in VM with slowly moving agents~\cite{holubec_finite-size_2021}, bringing its value closer to that observed in natural systems~\cite{cavagna2018physics}. However, the original VM appears to exhibit critical behavior only up to a certain system size, which scales inversely with the agents' speed. Beyond this threshold, the order–disorder transition takes on the character of a liquid–gas phase separation and appears discontinuous in finite-size systems~\cite{Solon2013,Chate2020}. It is accompanied by microscopic phase separation into dense bands that travel across the system~\cite{chate_collective_2008,Chate2020}. Whether the delayed VM exhibits similar behavior remains an open question.

In this work, we demonstrate that the order–disorder transition in the delayed VM \textcolor{black}{at a large but fixed system size} exhibits the same phenomenology as that in the standard VM. The phase retains the character of a liquid–gas phase separation~\cite{Solon2013,Chate2020}, marked by bistability, a negative Binder cumulant, and phase separation into dense, propagating bands. As the delay increases, the number of bands grows, while the time required for their formation decreases. This acceleration is driven by the emergence of swirls within the system, whose characteristic radius expands with increasing delay time. The relaxation time far away from the transition seems to be unaffected by the delay.

The rest of the paper is structured as follows. In the next Sec.~\ref{sec:model}, we describe the model and the parameter regime investigated. Our main results are given in Secs.~\ref{sec:order_param}--\ref{sec:bands}. In Sec.~\ref{sec:order_param}, we study the behavior of average polarization, polarization fluctuation, and the Binder cumulant as functions of delay. In Sec.~\ref{sec:bistability_times}, we further examine the bistability of the system in the liquid-gas phase and the effects of time delay on the relaxation time of the system. In Sec.~\ref{sec:bands}, we discuss the effects of time delay on the band formation time and the number of bands. We conclude in Sec.~\ref{sec:Conclusion}. \textcolor{black}{\ref{appx:CL} and \ref{appx:C1} contain the technical details of the calculation of the correlation length and the directional autocorrelation, respectively.}

\section{Model}
\label{sec:model}

The original VM~\cite{vicsek_novel_1995-1,chate_modeling_2008} is a paradigmatic example of dry aligning active matter~\cite{Vicsek201271,bechinger2016active}. The model describes an ensemble of $N$ agents moving at a constant speed~$v_0$ in two spatial dimensions and in discrete time. The position of the $i$th agent evolves according to
\begin{equation} \label{eq:2}
\mathbf{r}_i(t+\Delta t) = \mathbf{r}_i(t) + \Delta t\, \mathbf{v}_i(t+\Delta t),
\end{equation}
where $\Delta t$ is the time step and $\mathbf{v}_i(t+1)$ is the velocity vector at time $t+1$. The model is commonly simulated with periodic boundary conditions~\cite{nagy_new_2007,giraldobarreto2025}. 

At each time step, each agent aligns its velocity with the average velocity of its neighbors located within a distance~$R$ from its current position. The aligned velocity is then perturbed by noise. Since processing information about neighbors in practice necessarily takes some time, the delayed VM~\cite{piwowarczyk_influence_2019, holubec_finite-size_2021, geis_signal_2022} modifies the alignment rule so that, at time $t$, the agent aligns its velocity according to the average orientation of the neighbors it perceived $\tau$ time steps earlier, i.e., at time $t - \tau \Delta t$. The dynamical equation for the velocity of agent~$i$ thus reads
\begin{equation} \label{eq:1}
\mathbf{v}_i(t+\Delta t) = v_0 \left( \mathcal{R}_\eta \circ \vartheta \right) \left( \mathbf{v}_i(t) + \sum_{j \in S_i(t - \tau \Delta t)} \mathbf{v}_j(t - \tau \Delta t) \right),
\end{equation}
where $S_i(t - \tau \Delta t)$ denotes the set of neighbors located within a radius~$R$ from agent~$i$ at time~$t - \tau \Delta t$. The operator $\mathcal{R}_\eta \circ \vartheta(\mathbf{v})$ normalizes the input vector~$\mathbf{v}$ and then rotates it by a random angle~$\vartheta$ uniformly distributed in the interval~$\eta \pi [-1,1]$.

The natural time and length units in the model are the discrete time step~$\Delta t$ and the interaction radius~$R$, both of which are set to unity in our simulations. Two physically relevant dimensionless parameters emerge from this choice: the fraction of the interaction radius traversed by an agent in a single time step, $v_0 \Delta t / R$, and the distance covered during the delay time, defined as $\bar{\tau} \equiv v_0 \Delta t \tau / R$, which we refer to as the reduced delay time. In our simulations, we adopt a relatively high speed of $v_0 = 0.5$~\cite{nagy_new_2007}, yielding $v_0 \Delta t / R = 1/2$ and thus $\bar{\tau} = \tau / 2$. \textcolor{black}{This contrasts with previous studies on the delayed VM, which have typically employed significantly lower speeds~\cite{piwowarczyk_influence_2019, holubec_finite-size_2021}, e.g., $v_0=0.05$ in \cite{holubec_finite-size_2021}.}

\textcolor{black}{In our simulations, we considered $N = 65536$ and $131072$ agents in a square box of side length $L = 256$, with periodic boundary conditions. These setups correspond to mean agent densities of $\rho = N/L^2 = 1$ and $2$, respectively. Initially, at time $t = 0$, the agents were uniformly distributed throughout the simulation box and were assigned random orientations drawn from a uniform distribution. We assumed that the system was in this same state for all negative times as well, as required to iteratively solve the difference Eq.~\eqref{eq:1}. We then iterated Eqs.~\eqref{eq:2} and~\eqref{eq:1} simultaneously for all agents over $t = 10^{6}$ time steps. Our aim is to investigate the dynamical phases emerging in the system of given size $L$ and density $\rho$ as functions of the delay $\tau$ and noise intensity~$\eta$, after the system reaches a steady state independent of the initial conditions.
}

\section{Order parameters}
\label{sec:order_param}

The original VM exhibits three distinct phases governed by two primary control parameters: the particle density \(\rho\) and the noise intensity \(\eta\)~\cite{vicsek_novel_1995-1,solon_phase_2015}. At sufficiently high \(\rho\) and low \(\eta\), the agents spontaneously align their velocity vectors, breaking rotational symmetry~\cite{vicsek_novel_1995-1}. As \(\eta\) increases past the lower critical threshold \(\eta_{o|s}\), the system undergoes a liquid–gas phase transition characterized by dense, persistent traveling bands coexisting with a dilute disordered background~\cite{Solon2013,Chate2020,solon_phase_2015}. 
\textcolor{black}{This transition is only observed in systems above a minimum size—this threshold increases as particle speed decreases~\cite{chate_collective_2008}. Upon further increasing \(\eta\) beyond the upper critical threshold \(\eta_{s|d}\), the phase-separated state gives way to a fully disordered phase. In small systems, where band formation is suppressed, this order–disorder transition appears continuous. In larger systems, both transitions appear discontinuous due to the finite extent of the minimal nucleus of `liquid', and become continuous but non-critical in the infinite-size limit~\cite{Chate2020}.}

\textcolor{black}{A defining feature of the ordered states observed for noise levels below the upper critical threshold $\eta_{s|d}$ is polarization, defined as
\begin{equation} \label{eq:3}
\varphi(t) = \frac{1}{N} \left | \sum_{i=1}^N \frac{\mathbf{v}_i(t)}{|\mathbf{v}_i(t)|} \right|,
\end{equation}
which approaches its maximum value of one in a fully aligned system. Conversely, in the disordered states observed for $\eta > \eta_{s|d}$, the polarization remains close to zero.} The liquid–gas phase cannot be directly identified from the value of polarization; nevertheless, observing a bistable behavior when plotting polarization as a function of time points to this phase~\cite{Chate2020}. \textcolor{black}{However, the most reliable way to identify the liquid–gas phase is to combine the observation of bistability with additional indicators, such as the presence of bands in system snapshots. Beyond visual inspection, these bands manifest as a peak in the density autocorrelation function calculated along the direction of the system’s average polarization. The magnitude of this peak, $C_1$, therefore exhibits jumps at the phase boundaries of the liquid–gas phase and can be used to locate the transitions. For details, see~\ref{appx:C1} and Sec.~\ref{sec:bands}. For an alternative order parameter that can reveal the stripe phase, see Ref.~\cite{kursten_dry_2020}.}

Since polarization can exhibit significant fluctuations, we also measured the time-averaged polarization:
\begin{equation} \label{eq:4}
\langle \varphi \rangle = \frac{1}{t} \sum_{i=0}^{t/\Delta t} \varphi(i \Delta t),
\end{equation}
where the total simulation time \(t \) was sufficiently long to ensure that the initial equilibration period of the polarization did not affect the final value of the average.

Plotting the average polarization as a function of noise fails to capture the bistability evident in time series of the polarization. A more informative order parameter in this context is the Binder cumulant:
\begin{equation} \label{eq:6}
\langle U_4 \rangle = 1 - \frac{\langle \varphi^4 \rangle}{3\langle \varphi^2 \rangle^2}
\end{equation}
It exhibits a pronounced dip at the upper critical noise $\eta_{s|d}$, which marks the boundary between the disordered and bistable states; in sufficiently large systems, this dip can even become negative~\cite{chate_collective_2008}.

Besides the above order parameters, we also measured the scaled variance of the polarization:
\begin{equation} \label{eq:5}
    \langle\sigma^2\rangle = N(\langle\varphi^2\rangle - \langle\varphi\rangle^2),
\end{equation}
which serves as a measure of the system's susceptibility~\cite{bauerle_formation_2020}. In thermodynamic equilibrium, this definition of susceptibility agrees with an alternative definition based on correlation functions~\cite{cavagna2018physics}, which was applied to the delayed VM with small speed \(v_0\) in Ref.~\cite{holubec_finite-size_2021}. In that study, increasing delay was shown to cause the susceptibility to develop a tail, connected to the main peak by a point with a sharp change in its derivative.

\begin{figure}[h!]	
\centering
	 {\includegraphics[width=0.90
	\columnwidth]{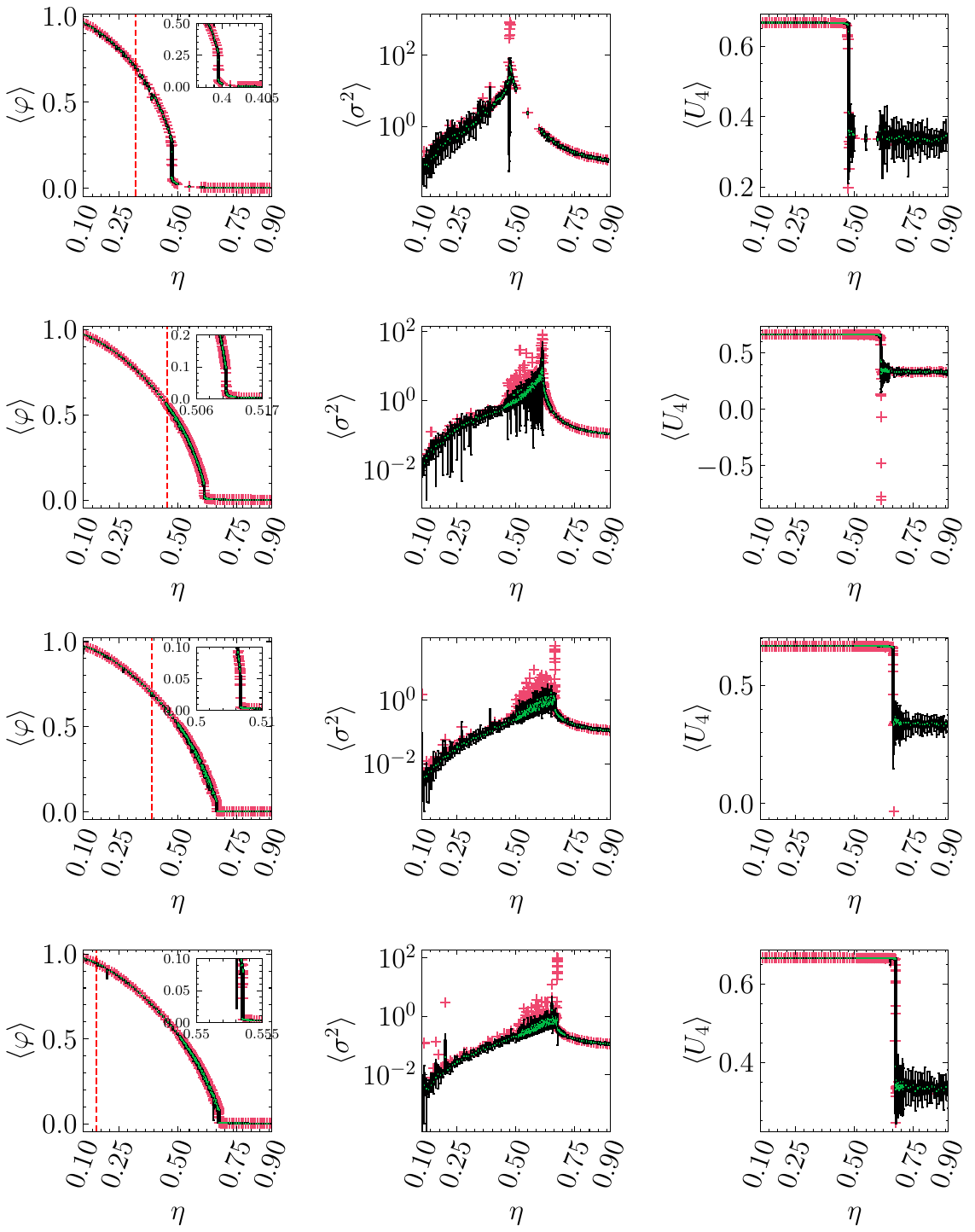}}
	\caption{\textbf{Order parameters vs. noise for $\rho =2$:} Average polarization $\langle \varphi \rangle$ (left), polarization variance $\langle \sigma^2 \rangle$ (middle), and Binder cumulant $\langle U_4 \rangle$ (right) as functions of noise intensity $\eta$ for reduced delay times $\tilde{\tau} = 0, 1/2, 3/2$, and $5/2$ (rows). Vertical lines in the first column show positions of the first, and insets magnify the region near the second transition. Boxplots display the distribution of the observable across 30 trajectory segments: green lines indicate the median, and black lines span the interquartile range. \textcolor{black}{They are longest near the second transition, where the system experiences the strongest fluctuations. Red crosses were obtained by averaging over entire trajectories, and their large spread is caused by outliers not shown in boxplots.}
  }
	\label{fig:1}	
\end{figure}

\begin{figure}
    \centering
    \includegraphics[width=1.0\columnwidth]{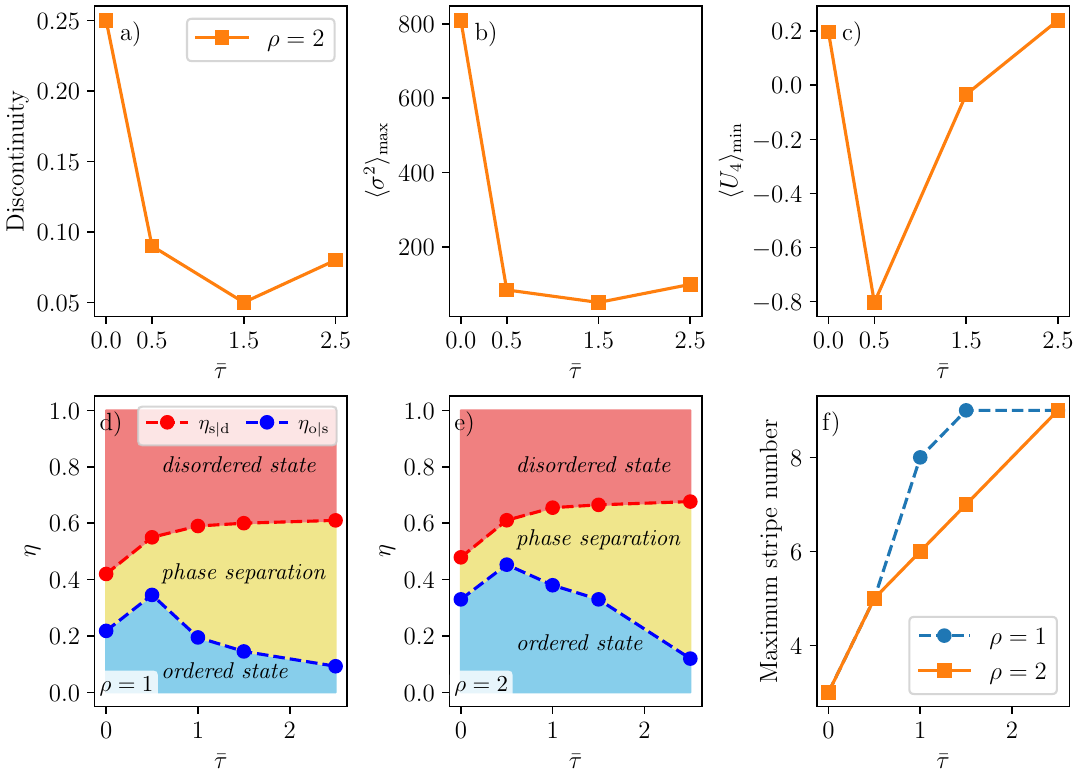}
	\caption{\textcolor{black}{\textbf{a)--c) Order parameters at the transition to disordered state ($\eta_{s|d}$) in Fig.~\ref{fig:1} ($\rho = 2$):}  
a) Magnitude of the jump in polarization at the transition.  
b) Polarization variance \(\langle \sigma^2 \rangle_{\mathrm{max}}\), which develops a maximum at the transition.  
c) Binder cumulant \(\langle U_4 \rangle_{\mathrm{min}}\), which develops a minimum at the transition. \textbf{d)–e) Phase diagrams for $\rho = 1$ and $2$, respectively:}  Dependence of the lower phase boundary \(\eta_{o|s}\), which separates the ordered state from the phase-separated state, and the upper phase boundary \(\eta_{s|d}\), which separates the phase-separated state from the disordered state, on the reduced delay \(\bar{\tau}\) and noise \(\eta\). \textbf{f) The maximum number of stripes} visually observed in the phase‑separated state for given $\tilde{\tau}$ and $\rho$ (for example snapshots of the system, see Fig.~\ref{fig:6}).}}
	\label{fig:2}	
\end{figure}

\textcolor{black}{In Fig.~\ref{fig:1}, we show the averaged order parameters \eqref{eq:4}--\eqref{eq:5} for $\rho = 2$ as functions of the noise intensity $\eta$, as obtained from our simulations. The results for $\rho = 1$ are qualitatively similar; however, both transitions are shifted to lower noise values, since reducing the density weakens the alignment interactions and thus lowers the system’s ability to form an ordered state.} Key observations are
\begin{enumerate}
    \item The average polarization \( \langle \varphi \rangle \) exhibits a discontinuity at the critical noise intensity \( \eta_{s|d} \) for all values of reduced delay time $\bar
    {\tau}$, showing that the transition into the disordered phase is discontinuous also in the delayed VM with a large but finite \( N \).
    \item The Binder cumulant \( \langle U_4 \rangle \) develops a pronounced dip at \( \eta_{s|d} \), revealing the presence of bistability for all values of reduced delay time $\bar
    {\tau}$.
    \item The magnitude of polarization variance $\langle \sigma^2 \rangle$ decreases with reduced delay time $\bar
    {\tau}$, and the sharp peak it develops for zero delay disappears with increasing $\bar{\tau}$.
    \item The position of the transition, $\eta_{s|d}$, shifts with increasing delay towards larger noise intensities.
    \item The averaged order parameters (\( \langle \varphi \rangle \)), \( \langle U_4 \rangle \), and \( \langle \sigma^2 \rangle \) do not provide insight into the phase-separated state and its transition to the ordered phase. \textcolor{black}{This insight can be obtained either from snapshots of the system at various noise values (see Fig.~\ref{fig:6}) or from the height of the first peak of the directional autocorrelation function, $C_1$ (see Figs.~\ref{fig:8}, \ref{fig:9}, and \ref{appx:C1}), both of which reveal the phase boundaries \( \eta_{o|s} \), indicated in the figure by the vertical lines in the first column.}
\end{enumerate}
\textcolor{black}{For $\rho = 2$, these observations are confirmed and further quantified in Fig.~\ref{fig:2}a--c, which show the most prominent features of the order parameters from Fig.~\ref{fig:1} at the transition as functions of the delay: the magnitude of the discontinuity in the average polarization, the maximum susceptibility, and the minimum Binder cumulant. Importantly, all these features saturate with increasing reduced delay time $\overline{\tau}$~\cite{holubec_finite-size_2021}. Moreover, Figs.~\ref{fig:2}e and~\ref{fig:2}f display, for both investigated densities $\rho = 1$ and $2$, the noise intensities at the transitions between the phase‑separated and disordered states, $\eta_{s|d}$, and between the ordered and phase‑separated states, $\eta_{o|s}$. The dependence of the upper phase boundaries, $\eta_{s|d}$, on the delay is monotonic and saturates with $\tilde{\tau}$. In contrast, the dependence of the lower phase boundaries, $\eta_{o|s}$, on the delay is non‑monotonic, and it remains unclear whether it saturates with $\overline{\tau}$ at a nontrivial noise value.
The width of the noise interval in which the liquid–gas coexistence state is observable clearly increases with the delay time.}

Besides the finding that the transition to the disordered state is discontinuous and accompanied by bistability regardless of the delay, the most striking observation is the enhanced stability of the phase‐separated state with nonzero polarization as the delay increases, manifested by a shift of the transition \(\eta_{s|d}\) toward higher noise intensities. Even at delays exceeding the characteristic interaction time—when agents detach from their neighbors before reengaging—\(\eta_{s|d}\) continues to saturate rather than decrease~\cite{holubec_finite-size_2021}, consistent with the plateauing of the order parameters in Fig.~\ref{fig:2}. While a future downturn of \(\eta_{s|d}\) at longer delays cannot be entirely excluded, such an outcome appears unlikely. \textcolor{black}{Also notable is the nonmonotonic behavior of the lower boundary $\eta_{o|s}$: it increases for short delays and decreases for longer ones, reflecting the competing effects of delay on phase separation described below.} Thus, time delay serves as a tunable control parameter, reversibly driving the system between the phase-separated and ordered states.

This behavior should be compared with the findings of Refs.~\cite{piwowarczyk_influence_2019,holubec_finite-size_2021} for the low-speed regime, where an initial increase in stability at short delays gives way to decreased stability at longer delays. However, in these studies, the phase-separated state was not observed. One could therefore speculate that the observed nonmonotonic role of delay on the stability of the ordered phase is analogous to the behavior of \(\eta_{o|s}\) observed here.

In Ref.~\cite{holubec_finite-size_2021}, the authors attributed the increased stability at short delays, followed by a decrease at longer delays, to the behavior of susceptibility, which quantifies the system’s responsiveness to external stimuli. High susceptibility enhances alignment among agents but also increases sensitivity to random fluctuations; as it decreases, alignment weakens while resistance to noise strengthens. This trade-off might explain the nonmonotonic dependence of stability on delay time observed.

\textcolor{black}{In the present high-speed regime, the susceptibility—quantified by the polarization variance \(\langle\sigma^2\rangle\)—drops sharply as \(\bar{\tau}\) approaches \(1/2\) and thereafter approximately saturates at a low value, similarly to the susceptibility reported in Ref.~\cite{holubec_finite-size_2021} for the low-speed regime. Hence, the qualitative change in the monotonicity of the stability increase should be attributed to speed.}

\textcolor{black}{Indeed, it is well established that the stability of the VM increases with particle speed. This conclusion is supported by the upward shift of the critical noise intensity with speed observed in previous studies~\cite{holubec_finite-size_2021,chate_collective_2008,vicsek_novel_1995-1} and by the correspondence between the zero-speed VM and the two-dimensional Heisenberg model, which cannot sustain long-range order due to the Mermin–Wagner theorem, whereas finite speed overcomes this limitation~\cite{ginelli_physics_2016}.}
Intuitively, nonzero speed stabilizes the system by enabling mobile “spins” to transport and dissipate fluctuations across the system. Time delay plays an analogous role by permitting agents to effectively average over temporal fluctuations. \textcolor{black}{Importantly, analytical results further indicate that, for sufficiently long delays, the effective control parameter in the delayed VM becomes \(v_0\,\tau\)~\cite{holubec_finite-size_2021}, and thus the observed increase in stability compared to the low-speed regime can indeed be explained by the large speed. To conclude, our findings for \(\eta_{s|d}\) suggest that, at high speeds, the reduction in susceptibility shifts the balance between alignment efficiency and noise resistance in favor of greater overall stability.
}

\begin{figure}	
\centering
\begin{tikzpicture}
	\node (img1)  {\includegraphics[width=1
	\columnwidth]{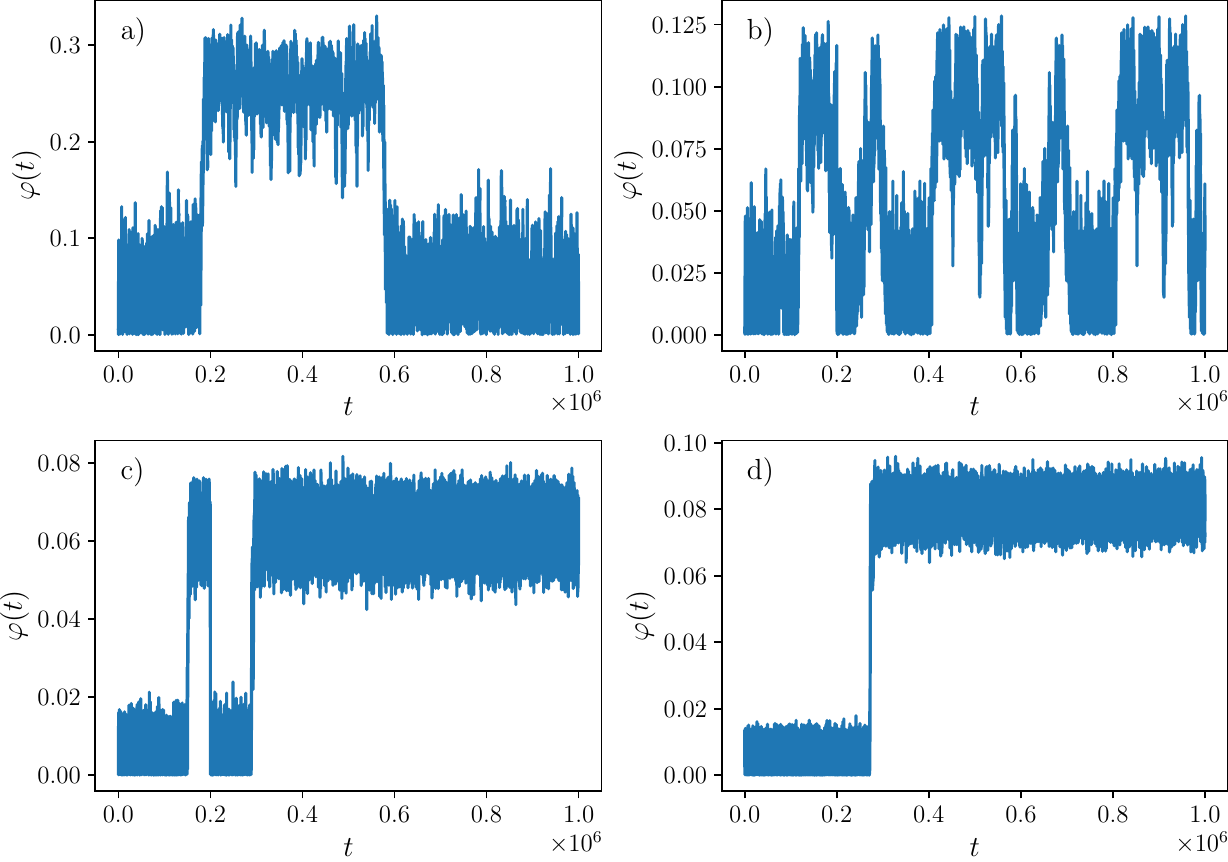}};
\end{tikzpicture}
	\caption{\textbf{Example time series at the transition for $\rho=2$:} a)–d) Time series of the average polarization at the order–disorder transition for $\bar{\tau} = 0$, $1/2$, $3/2$, and $5/2$, corresponding to critical noise intensities $\eta_{s|d} \approx 0.478$, $0.613$, $0.666$, and $0.677$, respectively.
    }
	\label{fig:3}	
\end{figure}

\begin{figure}
\centering
\begin{tikzpicture}
	\node (img1)  {\includegraphics[width=1.0
	\columnwidth]{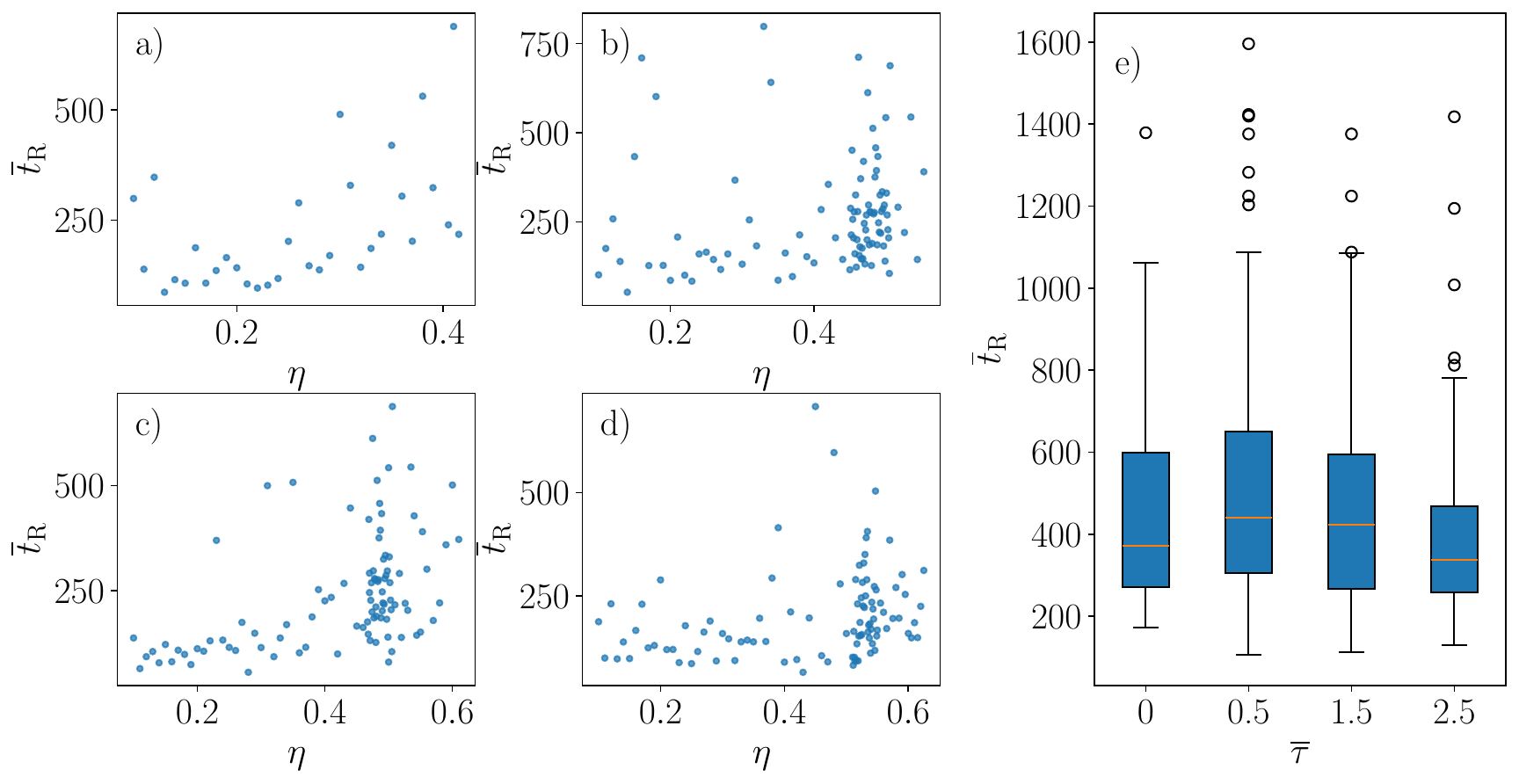}};
\end{tikzpicture}
	\caption{\textbf{Relaxation times for $\rho=2$:} a)–d) Reduced relaxation times $\bar{t}_{\rm R} = v_0 t_{\rm R}/R$, defined as the time required to reach half of the maximum polarization for the first time, plotted as functions of noise intensity $\eta$ for $\bar{\tau} = 0$, $1/2$, $3/2$, and $5/2$. e) Corresponding boxplots of the relaxation times as a function of $\bar{\tau}$. The orange dashed lines indicate the medians, the boxes represent the interquartile range (IQR), the whiskers extend to the most extreme data points within 1.5 times the IQR from the quartiles, and the circles denote individual outliers beyond this range.}
	\label{fig:4}	
\end{figure}

\section{Bistability and relaxation times}
\label{sec:bistability_times}

As discussed above, near the phase transition, the original VM exhibits bistability, manifested as a dip in the Binder cumulant and most clearly observed in individual polarization trajectories, which resemble Brownian motion in a double-well potential~\cite{holubec_finite-size_2021}. In our current simulations, the Binder cumulant similarly displays a dip across all studied delay times, suggesting that the corresponding time series likewise resemble bistable Brownian motion.

\textcolor{black}{This behavior is confirmed in Fig.~\ref{fig:3}, which shows time series of the polarization $\varphi(t)$ at the transition to the disordered state for reduced delay times $\bar{\tau} = 0, 1/2, 3/2,$ and $5/2$ at density $\rho = 2$.  
The corresponding time series for $\rho = 1$ and other reduced delay times $\bar{\tau}$ exhibit qualitatively the same behavior.} Interestingly, the presented series suggest that the transition rates between the more and less ordered states of polarization initially increase and then decrease with increasing $\bar{\tau}$. Nevertheless, since obtaining each time series required a considerable amount of computational time, we have not been able to further corroborate this hypothesis.

All available time series were used to measure the polarization relaxation time, $t_{\rm R}$, defined as the time at which $\varphi(t)$ first reaches half of its maximum value, i.e., $\varphi(t_{\rm R}) = \frac{1}{2} \max_{t} \varphi(t)$. \textcolor{black}{The results for reduced delay times $\bar{\tau} = 0, 1/2, 3/2,$ and $5/2$ at density $\rho = 2$ are shown in Figs.~\ref{fig:4}a)--d), with a summary boxplot presented in panel e). These results consistently indicate that delay has no significant effect on the relaxation time $t_{\rm R}$. Similar results were obtained for $\rho = 1$ and other delay values.}

Interestingly, the observation that the relaxation time of polarization, $t_{\rm R}$, remains nearly independent of the delay time, combined with the apparently nonmonotonic behavior of transition rates, and the findings of Ref.~\cite{holubec_finite-size_2021}—which demonstrate that the decay time of velocity correlations increases monotonically with delay—collectively underscore the intricate influence of time delay on the VM. This complexity becomes even more evident when examining the speed of band formation as a function of delay time, as explored in the subsequent section.

\section{Band formation and structure}
\label{sec:bands}

\begin{figure}[h!]	
\centering
\begin{tikzpicture}
	\node (img1)  {\includegraphics[width=0.95
	\columnwidth]{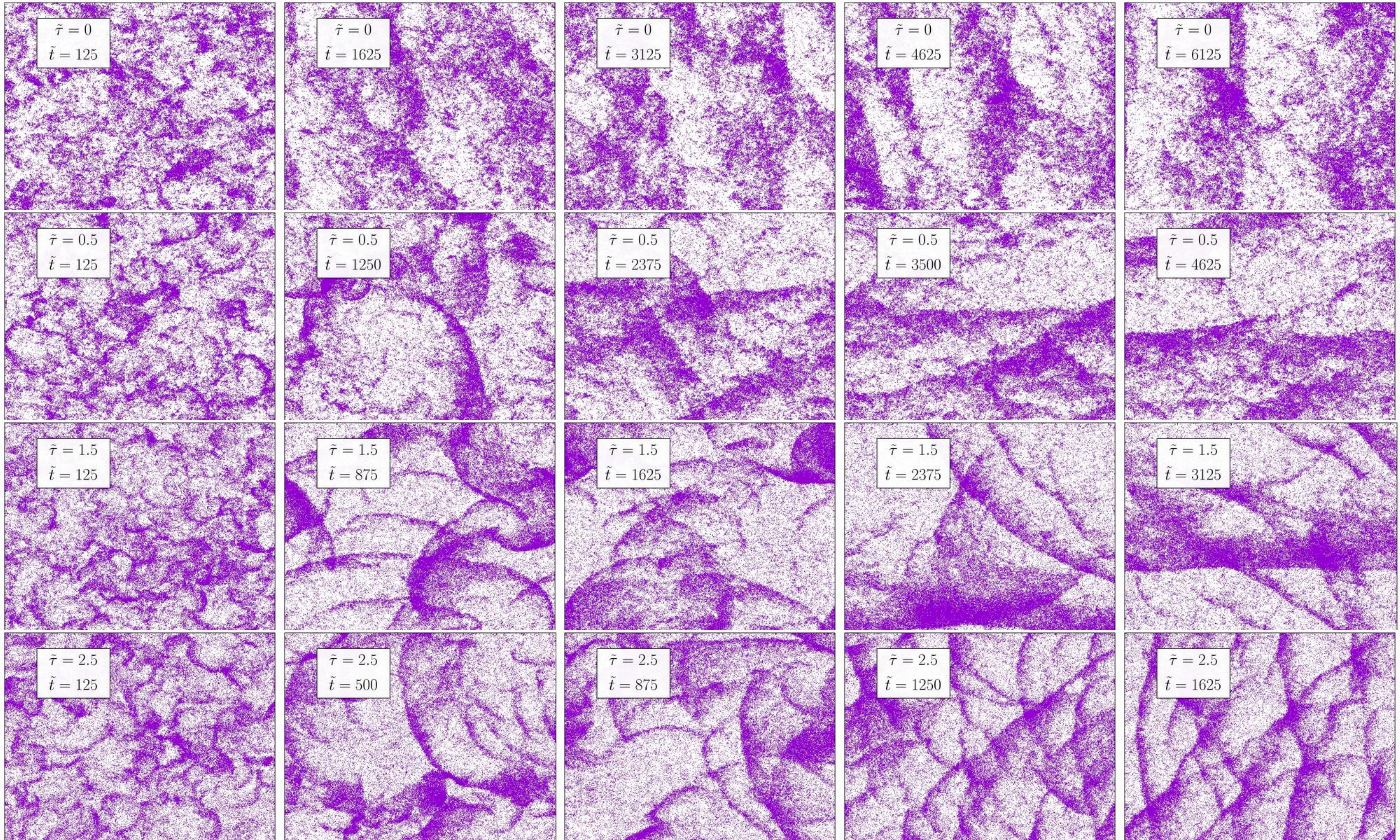}};
\end{tikzpicture}
	\caption{\textbf{Development of the bands for $\rho=2$:} Rows correspond to delay times $\bar{\tau} = 0$, $1/2$, $3/2$, and $5/2$, with associated noise values $\eta \approx \eta_{s|d} - 0.12 = 0.35$, $0.5$, $0.55$, and $0.57$, respectively. The snapshot in the $(n\!+\!1)$st column was taken at timestep $250(1 + n\Delta)$, with $\Delta = 12$, $9$, $6$, and $3$ for $\bar{\tau} = 0$, $1/2$, $3/2$, and $5/2$, respectively. 
    }
	\label{fig:5}	
\end{figure}

\begin{figure}[t!]	
\centering
\begin{tikzpicture}
	\node (img1)  {\includegraphics[width=1
	\columnwidth]{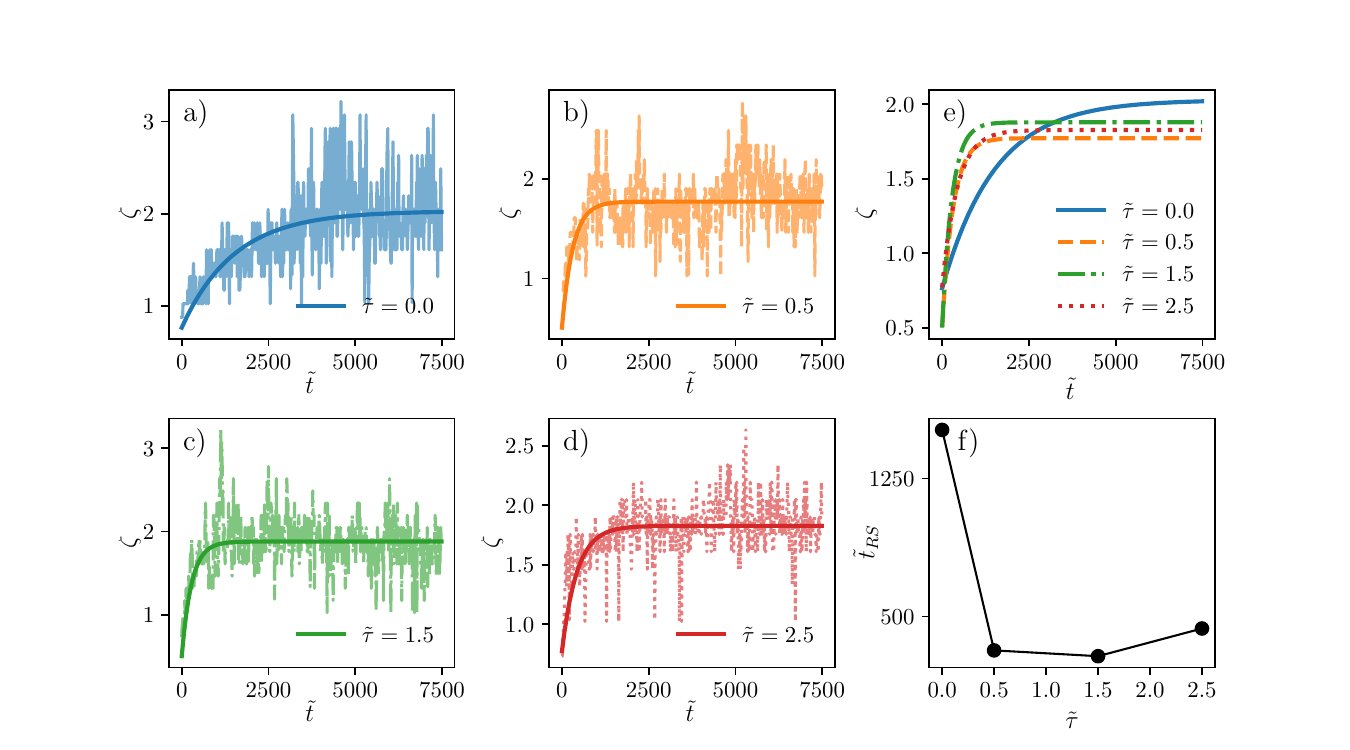}};
\end{tikzpicture}
	\caption{\textcolor{black}{\textbf{Relaxation of radial density correlation length for $\rho=2$:}  
a)--d) Radial density correlation length $\zeta$ as function of reduced time $\tilde{t}$ for reduced delay times $\tilde{\tau} = 0, 1/2, 3/2$, and $5/2$ together with fits with Eq.~\eqref{eq:fitzeta} (solid lines). Corresponding snapshots are shown in Fig.~\ref{fig:5}.  
e) Fits shown in panels a)--d).  
f) The reduced decay times $\tilde{t}_{RS}$ extracted from the fits in panels a)--d) as function of reduced delay time $\tilde{\tau}$.
}}
	\label{fig:7}	
\end{figure}

\textcolor{black}{Let us now examine the emergence of the phase‑separated state. In Fig.~\ref{fig:5}, we present snapshots from simulation intervals in which visible bands form for delay times $\bar{\tau} = 0,\;1/2,\;3/2,$ and $5/2$ and $\rho = 2$. Although the relaxation time in Fig.~\ref{fig:4} shows little dependence on $\bar{\tau}$, Fig.~\ref{fig:5} suggests that the time required to develop a discernible band structure decreases markedly with increasing delay (note that the interval between successive snapshots shrinks as $\bar{\tau}$ increases). To quantify the reduced stripe‑formation time \(\tilde{t}_{RS}\), we calculated the radial density correlation length \(\zeta\) (for details, see~\ref{appx:CL}) as a function of the reduced time \(\tilde{t} = v_0 t / R\), and we fitted the resulting curves with
\begin{equation}
a - b \exp\!\left(-\frac{\tilde{t}}{\tilde{t}_{RS}}\right).
\label{eq:fitzeta}
\end{equation}
The results plotted in Fig.~\ref{fig:7} indeed show that $\tilde{t}_{RS}$ decays with increasing delay until $\bar{\tau} = 3/2$ and then slightly increases.
} 

\textcolor{black}{The acceleration in stripe formation arises from the emergence of dense “arcs,” whose radii increase with $\bar{\tau}$, as can be seen in Fig.~\ref{fig:5}.} Long delays weaken collision‐induced interactions: agents in two approaching flocks only “sense” one another after they have begun to pass, so alignment occurs over a shortened window defined by their relative speed. In contrast, agents begin aligning for short delays while still in close proximity, allowing more time to coalesce into a single, well‐aligned flock. The reduced interaction time at long delays therefore yields a larger post‐collision spread in orientation, which manifests as arcs of increasing radius in Fig.~\ref{fig:5}.

\begin{figure}[t!]	
\centering
\begin{tikzpicture}
	\node (img1)  {\includegraphics[width=1
	\columnwidth]{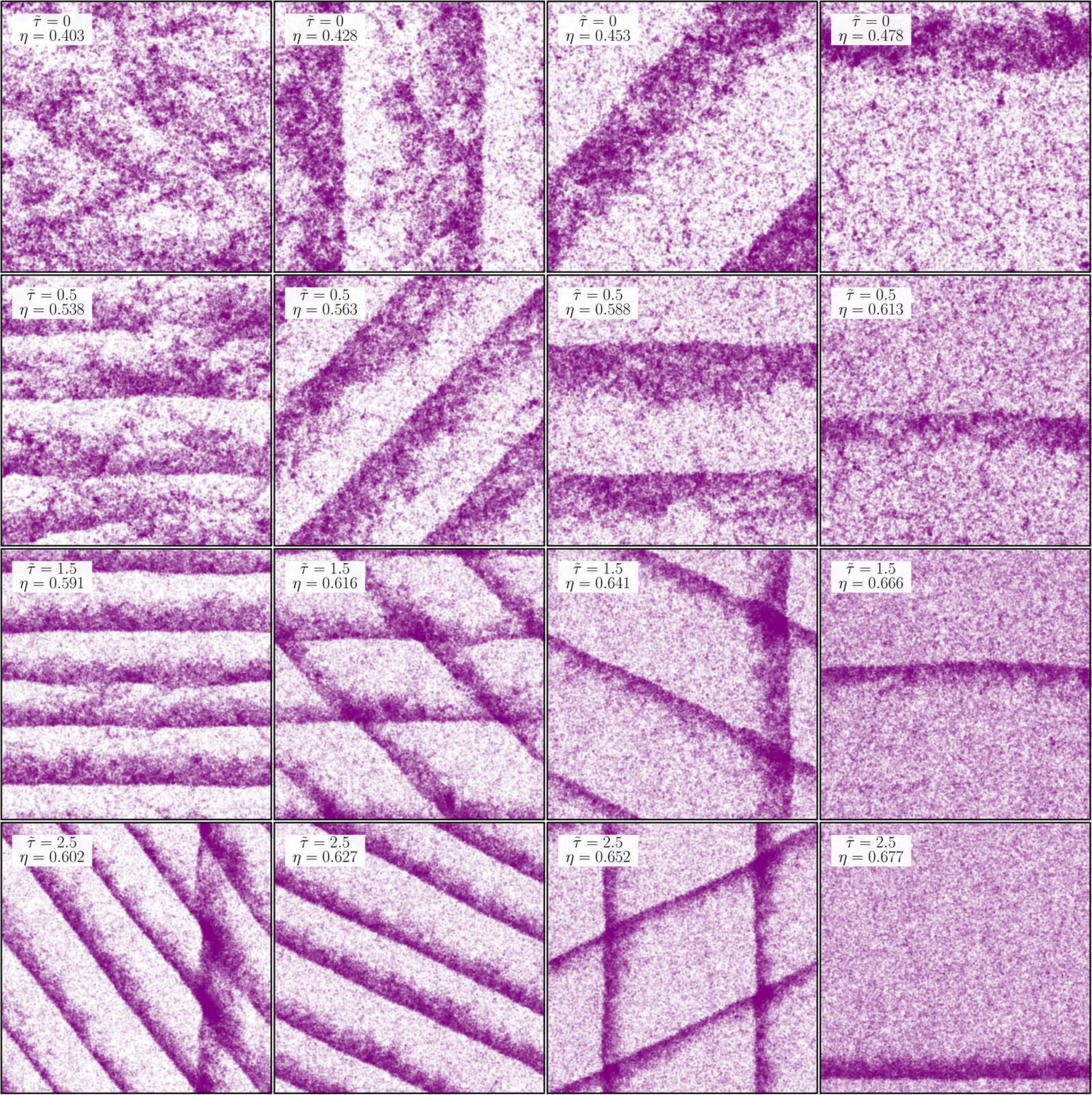}};
\end{tikzpicture}
	\caption{\textbf{Traveling bands for $\rho=2$:} Rows correspond to delay times $\bar{\tau} = 0$, $1/2$, $3/2$, and $5/2$ (from top to bottom), and columns correspond to noise values $\eta = \eta_{s|d} - n \cdot 0.5$ with $n = 0$, $0.05$, $0.1$, and $0.15$ (from right to left), where $\eta_{s|d}$ denotes the location of the transition to the disordered state ($\eta_{s|d} \approx 0.478$, $0.613$, $0.666$, and $0.677$ for $\bar{\tau} = 0$, $1/2$, $3/2$, and $5/2$, respectively). The number of stripes increases with $|\eta - \eta_{s|d}| \sim n$, up to a delay-dependent threshold $\eta_{o|s}$ ($\eta_{o|s} \approx 0.33$, $0.46$, $0.4$, and $0.16$ for $\bar{\tau} = 0$, $1/2$, $3/2$, and $5/2$, respectively), beyond which stripes no longer form and the system becomes ordered (cf.\ Fig.~\ref{fig:2}a). The stripes move in the direction of their denser fronts, as indicated by the black arrows in subplot 3 of the last row. Their speed is comparable to $v_0$. \textcolor{black}{The maximum number of stripes observed for each individual delay time is shown in Fig.~\ref{fig:2}.}
}
	\label{fig:6}	
\end{figure}

\begin{figure}[t!]	
\centering
\begin{tikzpicture}
	\node (img1)  {\includegraphics[width=1
	\columnwidth]{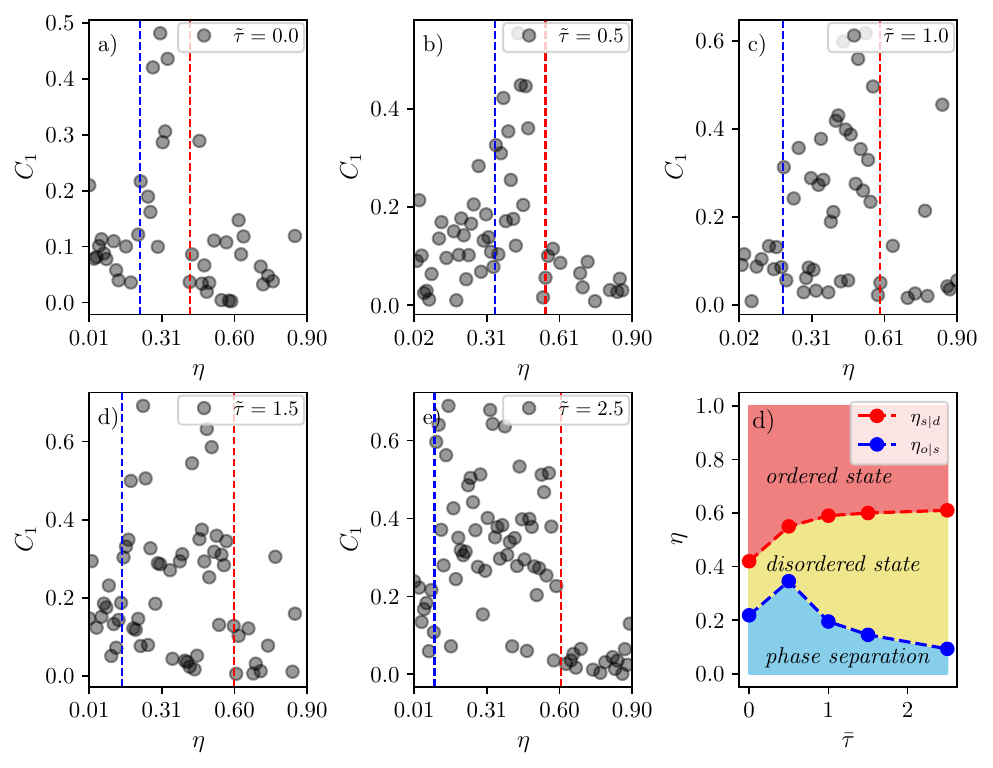}};
\end{tikzpicture}
	\caption{\textcolor{black}{\textbf{Phase boundary between ordered and phase-separated states for $\rho = 1$:}  
a)–e) Plots of the height $C_1$ of the first peak of the directional density autocorrelation function versus noise for reduced delay times $\tilde{\tau} = 0, 1/2, 1, 3/2,$ and $5/2$, respectively. The data exhibit a sharp increase at the transition noise intensity $\eta_{o|s}$, marked in the figures by blue dashed lines.  
The noise intensities $\eta_{s|d}$ at the transition between phase-separated and disordered states, marked in the figures by red dashed lines, were obtained from an analysis analogous to that in Fig.~\ref{fig:1}. They can, however, also be inferred from the \(C_1\) data as the noise intensities at which \(C_1\) decreases markedly.
Panel d) shows the resulting phase diagram, reproduced in Fig.~\ref{fig:2}d.}
}
	\label{fig:8}	
\end{figure}

\begin{figure}[t!]	
\centering
\begin{tikzpicture}
	\node (img1)  {\includegraphics[width=1
	\columnwidth]{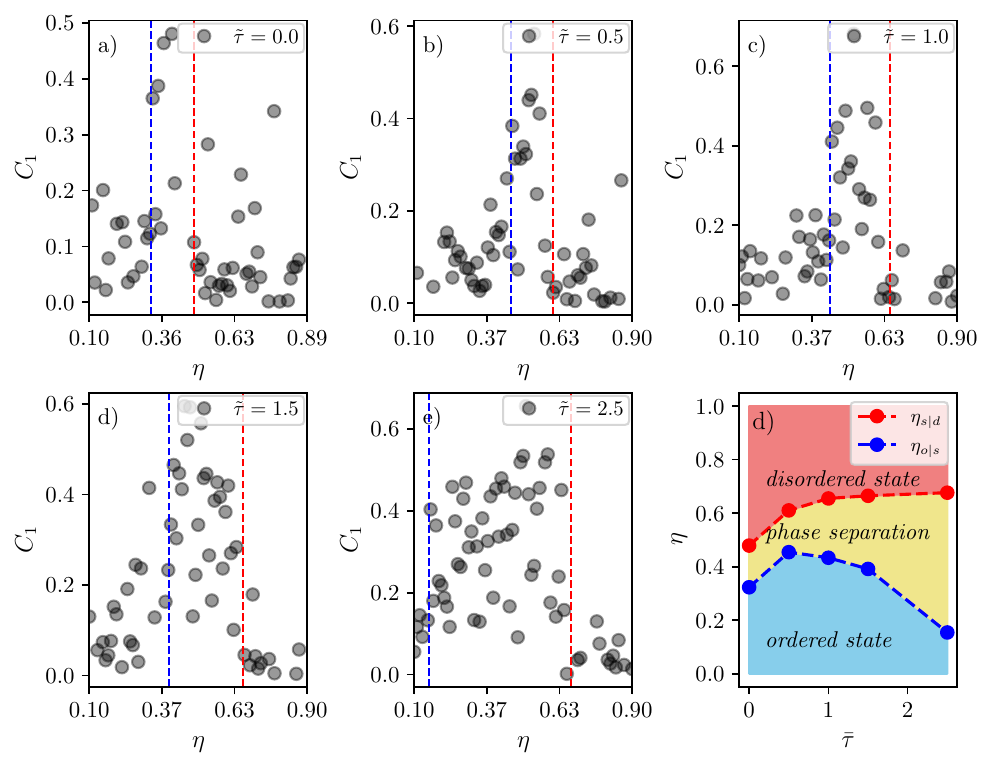}};
\end{tikzpicture}
	\caption{\textcolor{black}{\textbf{Phase boundary between ordered and phase-separated states for $\rho = 2$:} a)–e) Plots of the height $C_1$ of the first peak of the directional density autocorrelation function versus noise for reduced delay times $\tilde{\tau} = 0, 1/2, 1, 3/2,$ and $5/2$, respectively. The data exhibit a sharp increase at the transition noise intensity $\eta_{o|s}$, marked in the figures by blue dashed lines.  
The noise intensities $\eta_{s|d}$ at the transition between phase-separated and disordered states, marked in the figures by red dashed lines, were obtained from an analysis analogous to that in Fig.~\ref{fig:1}. They can, however, also be inferred from the \(C_1\) data as the noise intensities at which \(C_1\) decreases markedly.
Panel d) shows the resulting phase diagram, reproduced in Fig.~\ref{fig:2}e.}
}
	\label{fig:9}	
\end{figure}

\textcolor{black}{As shown in Figs.~\ref{fig:2}c and~d, the range of noise values over which phase separation—and thus the formation of bands—is observable increases with delay. Furthermore, the maximum number of observable stripes generally increases with delay time, as shown in Fig.~\ref{fig:2}e. Let us now provide numerical evidence supporting these conclusions. In Fig.~\ref{fig:6} we present snapshots of the system for $\rho = 2$, clearly showing that the number of bands grows with increasing delay. A qualitatively similar picture is observed for $\rho = 1$. The maximum stripe numbers plotted in Fig.~\ref{fig:2}c were obtained by visually analyzing all snapshots from simulations and determining the largest number of stripes observable for each delay time and density; consequently, the obtained numbers are subject to corresponding errors. To provide a more objective analysis, we plot in Figs.~\ref{fig:8} and~\ref{fig:9} the height of the first peak of the directional autocorrelation function as a function of noise $\eta$ for $\rho = 1$ and $2$, respectively. The peak height exhibits a sharp increase at the phase boundaries of the phase‑separated state, $\eta_{o|s}$ and $\eta_{s|d}$. These figures clearly show that the extent $\eta_{s|d} - \eta_{o|s}$ of the phase‑separated state widens with increasing delay. They also show that reducing the density shifts both transitions toward lower noise values, due to the reduced intensity of alignment interactions, as explained in Sec.~\ref{sec:order_param} above.
}

This is not entirely surprising, as the number of bands in the classical VM increases with agent speed $v_0$ \cite{chate_collective_2008, chate_modeling_2008, kursten_dry_2020}, and for long delays, the relevant parameter in the delayed VM becomes the product of speed and delay, as shown in the supplementary information of Ref.~\cite{holubec_finite-size_2021}. The number of stripes also increases with the distance from the phase transition, $\eta_{s|d} - \eta$. Consequently, with increasing delay and $\eta_{s|d} - \eta$, we observe more numerous bands. In some simulation runs, this led to the formation of so-called cross-see states, where stripes move in different directions~\cite{kursten_dry_2020}. However, in our simulations, we observe that for a given delay, cross-sea states appear at certain noise levels but not at lower noise levels (i.e., for larger $\eta_{s|d} - \eta$), even when the number of bands is comparable or higher. This raises uncertainty about whether the observed cross-sea configurations or the states with parallel bands are truly stationary. It remains unclear whether these patterns would eventually evolve into a single stable configuration or whether both the aligned and cross-sea states coexist as bistable solutions.

\section{Conclusion}
\label{sec:Conclusion}

In this study, we numerically investigated the impact of time delays on the phase behavior of the Vicsek model, a paradigmatic framework for studying collective motion in active matter systems. Our results, \textcolor{black}{obtained at a large but fixed system size}, demonstrate that the delayed Vicsek model retains the three characteristic phases of the standard model—ordered, liquid–gas coexistence, and disordered states—while the introduction of delay significantly alters the dynamics and stability of these phases.
Specifically, at the high agent speed considered ($v_{0}=0.5$), the critical noise intensity for the transition from the ordered to the coexistence state, $\eta_{o|s}$, exhibits a non-monotonic dependence on delay, initially increasing for short delays and decreasing for longer ones. In contrast, the critical noise for the transition to the disordered state, $\eta_{s|d}$, consistently increases with delay, broadening the noise interval where phase separation occurs.

The liquid–gas coexistence phase, marked by dense traveling bands, is notably stabilized by delay: the number of bands increases and their formation time decreases as delay grows. This acceleration is driven by the emergence of swirling structures, whose characteristic radius expands with delay, reflecting weakened collision-induced interactions at longer delays. The bistability of the system, evident in polarization time series and a negative Binder cumulant, persists across all delay times, confirming that the order–disorder transition retains its discontinuous nature in large but finite systems. Interestingly, while delay significantly influences phase boundaries and band dynamics, the relaxation time of polarization remains largely unaffected, underscoring the complex interplay between delay and system dynamics.

Our findings establish time delay as a powerful control parameter for tuning the dynamic phase behavior of the Vicsek model, offering insights into the role of delayed interactions in active matter systems. The non-monotonic effects of delay on phase stability and the enhanced stability of the phase-separated state at high speeds suggest analogies with natural systems, such as bacterial swarms or animal flocks, where perception–reaction lags are inherent. Future work could explore the thermodynamic limit to clarify the nature of the observed discontinuities \textcolor{black}{and the asymptotic values of $\eta_{o|s}$ and $\eta_{s|d}$}, investigate the stability of cross-sea states versus parallel-band configurations, and extend the model to incorporate additional realistic features, such as variable delays or environmental heterogeneity. These advancements would further bridge the gap between theoretical models and experimental observations of collective behavior in biological and artificial active matter systems.

\section*{Acknowledgments}
VH gratefully acknowledges Charles University's funding through Project PRIMUS/22/SCI/009. We also acknowledge Klaus Kroy's invaluable discussions and guidance.

\appendix

\textcolor{black}
{
\section{Radial density correlation length $\zeta$}
\label{appx:CL}
}

\textcolor{black}
{
To calculate the radial correlation length $\zeta$, we divided the square arena into $53 \times 53 = 2809$ square patches and evaluated the local densities in each patch, obtaining the density field $\rho(\mathbf{r})$.
Next, we introduced the density fluctuation field,
\begin{equation}
\tilde{\rho}(\mathbf{r}) = \rho(\mathbf{r}) - \langle \rho \rangle,
\end{equation}
with the average density $\langle \rho \rangle = \frac{1}{N} \sum_{\mathbf{r}} \rho(\mathbf{r}) = \rho$,
and defined the two-dimensional spatial autocorrelation function
\begin{equation}
C(\mathbf{r}) = \sum_{\mathbf{r}'} \tilde{\rho}(\mathbf{r}') \tilde{\rho}(\mathbf{r}' + \mathbf{r}).
\end{equation}
We then normalized it as $\tilde{C}(\mathbf{r}) = C(\mathbf{r}) / C(\mathbf{0})$, and computed the radially averaged correlation function:
\begin{equation}
g(r) = \langle \tilde{C}(\mathbf{r}) \rangle_{|\mathbf{r}| = r} = \frac{1}{N_r} \sum_{|\mathbf{r}| = r} \tilde{C}(\mathbf{r}),
\end{equation}
where $N_r = \sum_{|\mathbf{r}| = r} 1$ is the number of points at a distance $r$ from the origin.
Finally, we defined the correlation length $\zeta$ as the distance at which $g(r)$ first decays to $1/{\rm e}$:
\begin{equation}
g(\zeta) = \frac{1}{e}.
\end{equation}
We solved this equation by linearly interpolating between discrete radial distances to determine where $g(r)$ crosses the $1/{\rm e}$ threshold.
}

\textcolor{black}
{
\section{Directional autocorrelation peak height $C_1$}
\label{appx:C1}
}

\textcolor{black}
{
To identify the liquid--gas coexistence state, we calculated the height of the first positive peak \( C_1 \) in the autocorrelation function \( \tilde{C}_\parallel(r)  \) of the marginal density \( g_\parallel(r) \), projected along the direction of motion.}

\textcolor{black}
{
To compute the marginal density, we first evaluated the average polarization vector for a given system state at time \( t \) in the steady state:
\begin{equation} \label{eq:polarization}
\boldsymbol{\varphi} = \frac{1}{N} \sum_{i=1}^N \frac{\mathbf{v}_i}{|\mathbf{v}_i|}.
\end{equation}
Next, we computed the centered coordinates \( \tilde{\mathbf{r}}_i = \mathbf{r}_i - \langle \mathbf{r} \rangle \), where \(
\langle \mathbf{r} \rangle = \frac{1}{N} \sum_{i=1}^{N} \mathbf{r}_i\).
We then projected the centered positions onto the average polarization vector:
\begin{equation}
\bar{x}_i = \boldsymbol{\varphi} \cdot \tilde{\mathbf{r}}_i.
\end{equation}
Using these projected coordinates, we constructed the coarse-grained marginal density \( g_\parallel(r) \) as a histogram:
\begin{equation}
g_\parallel(r) = \sum_{i=1}^{N} I\left(\bar{x}_i \in \left[r - \frac{dr}{2}, r + \frac{dr}{2} \right]\right),
\end{equation}
where \( dr \) is the bin width and \( I \) is the indicator function. In our analysis, we used thirteen bins, i.e., \( dr = L_\parallel/13 \), where \( L_\parallel \) is the total range of \( \bar{x}_i \) values in the snapshot.}

\textcolor{black}
{
We then introduced the centered density:
\begin{equation}
\tilde{g}_\parallel(r) = g_\parallel(r) - \langle g_\parallel \rangle,
\end{equation}
with \( \langle g_\parallel \rangle \) being the average of \( g_\parallel(r) \) over all bins.
The one-dimensional density autocorrelation function was computed as:
\begin{equation}
C_\parallel(r) = \sum_{r'} \tilde{g}_\parallel(r') \tilde{g}_\parallel(r' + r).
\end{equation}
The correlation function was then normalized as:
\begin{equation}
\tilde{C}_\parallel(r) = \frac{C_\parallel(r)}{\max_{r \ge 0} |C_\parallel(r)|}.
\end{equation}
Finally, the height \( C_1 \) of the first positive peak of \( \tilde{C}_\parallel(r) \) (excluding the origin \( r = 0 \)) was extracted. If no such peak exists, or if the first positive peak lies at the boundary of the support of \( \tilde{C}_\parallel(r) \)—which may result from self-correlations due to periodic boundaries—no value is assigned to \( C_1 \). The data shown in Figs.~\ref{fig:8}	and \ref{fig:9}	 were averaged over ten processed snapshots.}

\section*{References}

\bibliographystyle{iopart-num}
\bibliography{Bibliography}


\end{document}